\documentclass[12pt]{iopart}

\usepackage[dvips]{graphicx}

\def \beq {\begin{equation}}
\def \eeq {\end{equation}}

\begin{document}
\title{Pesin-type relation  for subexponential instability}

\author{Alberto Saa}\ead{asaa@ime.unicamp.br}
\address{Departamento de Matem\'atica Aplicada, UNICAMP, 13083-859, Campinas, SP, Brazil}
\author{Roberto Venegeroles}\ead{roberto.venegeroles@ufabc.edu.br}
\address{Centro de Matem\'atica, Computa\c c\~ao e Cogni\c c\~ao, UFABC, 09210-170, Santo Andr\'e, SP, Brazil}

\begin{abstract}
We address here the problem of extending the Pesin relation among positive
 Lyapunov exponents and the Kolmogorov-Sinai entropy  to the case of
dynamical systems exhibiting subexponential  instabilities. By using a recent rigorous result   due to Zweim\"uller, we show that the usual Pesin relation can be extended straightforwardly for weakly chaotic one-dimensional systems of
 the Pomeau-Manneville type, provided one introduces a convenient subexponential generalization of the Kolmogorov-Sinai entropy.
We show, furthermore, that Zweim\"uller's result provides an efficient prescription
for the evaluation of the algorithm complexity for such systems.
 Our results are confirmed by exhaustive numerical simulations. We also point out and correct a  misleading extension of the Pesin relation based on the Krengel entropy that has appeared recently in the literature.
\end{abstract}


\maketitle

\section{Introduction}

One-dimensional chaotic dynamics  are usualy characterized by the existence of a positive Lyapunov exponent, which indicates exponential separation of initially nearby trajectories \cite{PG}.  In recent years, we have witnessed a rapid  development  in the study and characterization of dynamical unpredictable  systems in  which the separation of trajectories is weaker than exponential \cite{GW}. For these systems, generically dubbed weakly chaotic in the physical literature,
the conventional Lyapunov exponent  vanishes and new concepts and ideas for the characterization of dynamical instabilities are necessary for a deeper understanding  of their global dynamics. Many results of infinite ergodic theory \cite{Aaronson} come  out  as   powerful tools in this context, shedding  light on several apparently unrelated results in the physical literature. Among them, the Aaronson-Darling-Kac (ADK) theorem \cite{Aaronson} has certainly a central role in these problems.

The main purpose of this work is to extend the well-known Pesin relation \cite{Pesin} for the case of weakly chaotic one-dimensional systems, a matter that has attracted considerable attention recently (see \cite{KB} for references) and even some controversy \cite{GTA}. For usual one-dimensional chaotic systems,  the Pesin relation is given simply by  $h=\lambda$, with $h$ and $\lambda$ standing, respectively, for the Kolmogorov-Sinai (KS) entropy   and the usual Lyapunov exponent. We will show that adequate subexponential  generalizations of the KS entropy and of the Lyapunov exponent will obey exactly the same Pesin-type relation, for almost all trajectories. It is important to stress that  the existence of such  generalization is far from intuitive since we are dealing  with nonergodic systems for which the typical dynamical quantities do depend  on the specific
 trajectory.
We also show that the extension  based on Krengel entropy proposed in \cite{KB} for weakly chaotic systems is incorrect. The source of the problem is identified and the correct expression is presented. We close by comparing  our proposed Pesin-type relation  based on the subexponential KS entropy and the    proposal involving the Krengel entropy.

\section{Pesin-type relation and statistics}

We will consider here  the general class of Pomeau-Manneville (PM) \cite{PM} maps $x_{t+1}=f(x_{t})$,  with $f:[0,1]\to [0,1]$ such that
\begin{equation}
\label{pmap}
f(x)\sim x(1+ax^{z-1}),
\end{equation}
for $x \to 0$, with $a>0$ and $z>1$. From the physical point of view, the original
PM system ($z=2$) is paradigmatic since it corresponds to certain Poincar\'e sections related to the Lorenz attractor \cite{PM}. Systems of the type (\ref{pmap}) exhibit exactly the kind of subexponential instability for nearby trajectories that we are concerned here: $\delta x_{t}\sim\delta x_{0}\exp(\lambda_{\alpha} t^{\alpha})$, with $0<\alpha<1$. A distinctive  characteristic of such class of maps is the presence of an indifferent (neutral) fixed point at $x=0$, {\em i.e.}, $f(0)=0$ and $f'(0)=1$. The global form of $f$ is irrelevant for our purposes, provided it respects the axioms of an AFN-system \cite{RZ}. These systems are known to have    invariant densities
$\omega(x)$
such
that
\beq
\label{dens}
\omega(x)\sim  {b}{x^{-\frac{1}{\alpha}}}, \quad \alpha=(z-1)^{-1},
 \eeq
 near the indifferent fixed point $x=0$  \cite{Thaler}. Clearly, the corresponding invariant measures diverge for $z>2$.  In these cases,   we have pictorially two qualitative distinct dynamical behavior coexisting, namely a laminar regime near $x=0$  and a turbulent one elsewhere, resulting eventually in nonergodicty and subexponential separation of initially close trajectories. It is noteworthy here that  it was recently shown  that subexponencial instability does imply infinite invariant measure \cite{AAC}. On the other hand, $1<z<2$ leads to a finite invariant measure, which is naturally related to ergodicity and positivity of the ordinary Lyapunov exponent.

For intermittent systems like (\ref{pmap}), the statistics of a given observable $\vartheta$ for randomly distributed initial conditions  has some peculiar properties. For ergodic systems, the time average  $t^{-1}\sum_{k=0}^{t-1}\vartheta(f^{k}(x))$ converges to the spatial average $\int\vartheta\,d\mu$, with $d\mu = \omega(x) dx$. On the other hand, if the system has a diverging invariant measure, the time average will typically depend on the chosen trajectory. Nevertheless, the ADK theorem \cite{Aaronson} ensures in this case that a suitable time-weighted average does converge  in distribution terms towards a Mittag-Leffler distribution with unit first moment. More specifically, for a positive integrable function $\vartheta$ and a random variable $x$ with an absolutely continuous measure with respect to the Lebesgue measure on the interval of interest, there is a sequence $\left\{a_{t}\right\}$ for which
\beq
\frac{a_{t}^{-1}\sum_{k=0}^{t-1}\vartheta(f^{k}(x))}{\int\vartheta d\mu}
\stackrel{d}{\longrightarrow}\xi_{\alpha},
\eeq
for $t\to\infty$, where $\xi_{\alpha}$ is a non-negative Mittag-Leffler random variable with index $\alpha\in (0,1]$ and  unit expected value. Notice that for $1<z<2$ (the ergodic regime), $a_{t}\sim t$, and the corresponding $\alpha=1$ Mittag-Leffler distribution reduces to a Dirac $\delta$-function.
 For the subexponential regime ($z>2$), we have $a_{t}\sim t^\alpha$ \cite{RZ} and, by choosing $\vartheta=\ln|f'|$, the ADK theorem assures the  convergence in distribution terms towards a Mittag-Leffler distribution of the subexponential finite-time Lyapunov exponent
\begin{equation}
\lambda^{(\alpha)}_{t}(x)=\frac{1}{t^{\alpha}}\sum_{k=0}^{t-1}\ln\left|f'\left(f^{k}(x)\right)\right|,
\label{lamba}
\end{equation}
for $t\rightarrow\infty$.
The generalized  Lyapunov exponent (\ref{lamba}) plays for intermittent systems the same role did by the usual exponent  (corresponding to $\alpha=1$ in (\ref{lamba})) for one-dimensional chaotic systems, see \cite{PSV} and references therein for a recent discussion.

In order to investigate the connection  between subexponential instability and the corresponding degree of randomness of an intermittent dynamical system  like (\ref{pmap}), we will consider the Kolmogorov-Chaitin concept of complexity  \cite{PG}. Let us assume that the phase space of the map (\ref{pmap}) is partitioned and completely covered by a set of non overlapping   ordered cells. A given trajectory $\{x_{t}\}$ generated by the map (\ref{pmap}) can be represented by a sequence of symbols $\{s_{t}\}$, which we assume to be integers such that $s_t$ corresponds to the cell  where  $x_{t}$ belongs. The next step in the analysis consists in  eliminating redundancies that may appear in $\{s_{t}\}$ by performing a compression of information. This can be done, for instance, by introducing the so-called algorithmic complexity function $C_t(\{s_{t}\})$, which is defined as the  length of the shortest possible program able to reconstruct the sequence $\{s_{t}\}$ on a universal Turing machine \cite{PG}. Systems that exhibit some degree of regularity are able to generate sequences of symbols at a rate higher than needed for recording their programs. For example, a periodic sequence can be recreated by replaying the periodic pattern over the total lenght. Typically, for these cases, one has $C_{t}\sim \ln t$. On the other hand, if the trajectory is completely random, there is no way of reproducing it other than memorizing the whole trajectory, resulting in a sequence length that increases linearly in time, {\em i.e.}, $C_{t}\sim t$. The finite time KS entropy is defined simply as $h_t = C_t/t$. An important recent rigorous result due to Zweim\"uller \cite{BWZ} unveils the relation between KS entropy and the Lyapunov exponent for systems exhibiting  subexponential instability. According to this result, we have, for almost all initial conditions,
\begin{eqnarray}
\label{smb}
\frac{C_{t}}{\sum_{k=0}^{t-1}\vartheta(f^{k}(x))}\rightarrow\frac{h_{\mu}(f)}{\int\vartheta d\mu},
\end{eqnarray}
for $t\rightarrow\infty$, for any observable function $\vartheta$, where $h_{\mu}(f)$ stands for the Krengel  entropy \cite{Kr}, which can be expressed by the
  so-called  Rohlin's formula \cite{Rohlin}
 \beq
 \label{rohlin}
 h_{\mu}(f)=\int\ln|f'| d\mu.
 \eeq
 By choosing again $\vartheta=\ln|f'|$, we get from (\ref{smb}) the   surprisingly simple
 relation
\begin{eqnarray}
\label{pes}
h_{t}^{(\alpha)}\rightarrow\lambda_{t}^{(\alpha)},
\end{eqnarray}
for $t\to\infty$ and for almost all initial conditions, where
\begin{equation}
\label{KS}
h_{t}^{(\alpha)} = \frac{C_t}{t^\alpha}
\end{equation}
is the subexponential generalization of the finite-time KS entropy. The relation (\ref{pes}) is the most natural generalization of the Pesin relation for systems of the type (\ref{pmap}). From the ADK theorem and (\ref{pes}), we have that both $h_{t}^{(\alpha)}$ and $\lambda_{t}^{(\alpha)}$ converge in distribution terms
toward the same Mittag-Leffler distribution. Nevertheless, Zweim\"uller's result is indeed stronger, assuring that, for almost all trajectories, $h_{t}^{(\alpha)}$ coincides with $\lambda_{t}^{(\alpha)}$ in the limit $t\to\infty$. In this way, the relation (\ref{smb}) does provide  an efficient
 prescription for evaluating the algorithmic
complexity of a given trajectory for one-dimensional maps, namely
\begin{equation}
\label{K}
C_t \to \sum_{k=0}^{t-1}\ln\left|f'\left(f^k(x)\right)\right|,
\label{ac}
\end{equation}
for large $t$. The power of the prescription (\ref{K}) resides in the fact that it does provide, for the systems in question, a computable way for the calculation of the algorithmic complexity function $C_t$, a well-known non-computable function in general \cite{PG}. It is important also to remind that, for dynamical systems with infinite invariant measure, the invariant density, and consequently, the invariant measure, is defined up to an arbitrary multiplicative positive constant. In other words, the transformation $\omega\to\xi\omega$ (implying, in this way, that $b\to\xi b$ in (\ref{dens})), with $\xi>0$, does not have any dynamical implication. Zweim\"uller's construction,
based in (\ref{smb}), is clear invariant under such transformation.
 Of course, such ``symmetry''
is broken in the usual ergodic case due to the normalization of the invariant measure.

We notice that the relation (\ref{pes}) is compatible with the pioneering work of Gaspard and Wang \cite{GW},  where the standard PM map $f(x)=x+ax^{z}$ (mod 1) was considered. They argue, in particular, that the algorithmic complexity $C_t$ for the PM map is proportional to $N_t$, the number of entrances into a given phase
space cell during $t$ iterations of the PM map. By invoking some results from renewal theory \cite{Feller}, one has
\begin{eqnarray}
\label{ren}
{\rm Prob}\left(N_{t}\geq c\frac{t^{\alpha}}{u^{\alpha}}\right)\rightarrow G_{\alpha}(u),
\end{eqnarray}
for $0<\alpha<1$ and $t\rightarrow\infty$, where $c$ is a positive constant  and $G_{\alpha}$ stands for the one-sided L\'evy cumulative distribution function with index $\alpha$. With the change of variable $u=r\xi^{-1/\alpha}$, where $r^{\alpha}=\alpha\Gamma(\alpha)$, we have that the  normalized random variable $\xi=N_{t}/\left\langle N_t\right\rangle$ tends toward a Mittag-Leffler random variable with index $\alpha$ and unit first moment for $t\to\infty$ (see \cite{SV} for more details on the relations between one-sided L\'evy and Mittag-Leffler distributions), in perfect agreement with the predictions of the ADK theorem.
The possibility of estimating the algorithmic complexity
function $C_t$ from $N_t$ also
for generic systems of the type (\ref{pmap}) is  indeed confirmed by
our exhaustive numerical explorations. (See, also, \cite{RZ} and references therein.)

\section{Numerical simulations}

\begin{figure}[t]
\begin{flushright}
\includegraphics[width=0.85\linewidth]{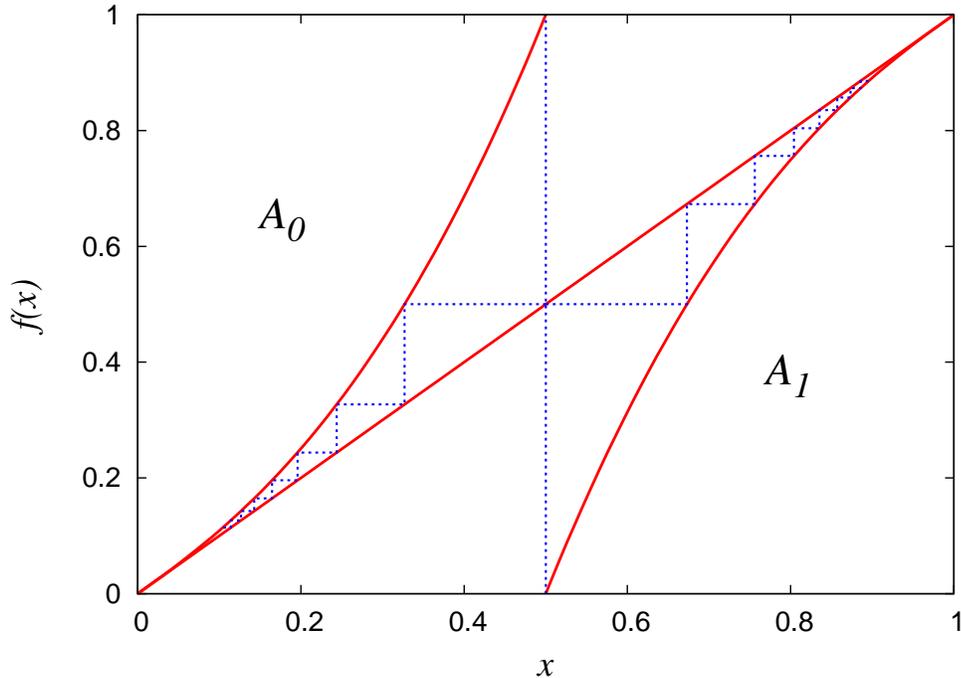}
\end{flushright}
\caption{The cells $A_0$ and $A_1$ for the Bernoulli map (\ref{bern}) with $z=5/2$. Notice that the   dynamics are regular inside each of the cells, with the trajectories departing monotonically  from the respective fixed points. Nevertheless, the transition for one cell to the other is chaotic. The situation is similar for the PM map with $a=1$ and for the Thaler map (\ref{tmap}), even though for theses cases the partitions are not symmetric as the present case. (See also \cite{GW}).}
\label{fig0}
\end{figure}

\begin{figure}[t]
\begin{flushright}
\includegraphics[width=0.85\linewidth]{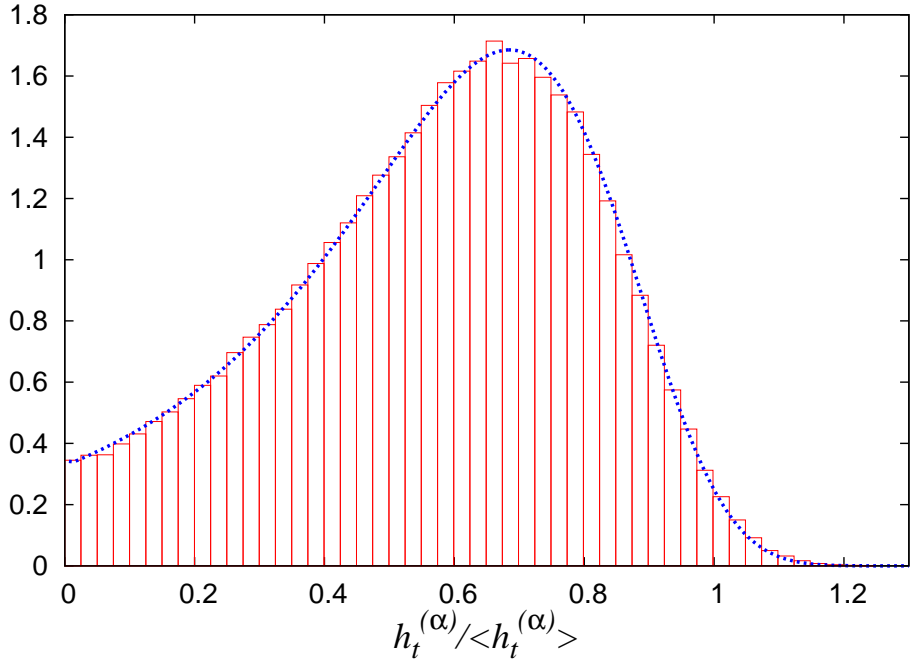}
\end{flushright}
\caption{Distribution of finite time Kolmogorov-Sinal entropy $h^{(\alpha)}_t$ calculated from (\ref{KS1}) for the Bernoulli map (\ref{bern}), with $z=28/13$ ($\alpha = 13/15$), for $t=6\times 10^4$ and $2.5\times 10^5$ uniformly distributed initial conditions. The histogram was built directly from the numerical data, while the dotted  line is the corresponding Mittag-Leffler probability density  computed with the algorithm of   \cite{SV}. For convergence details, see \cite{PSV}. Similar results hold also for
the other considered maps.}
\label{fig1}
\end{figure}

\begin{figure}[t]
\begin{flushright}
\includegraphics[width=0.85\linewidth]{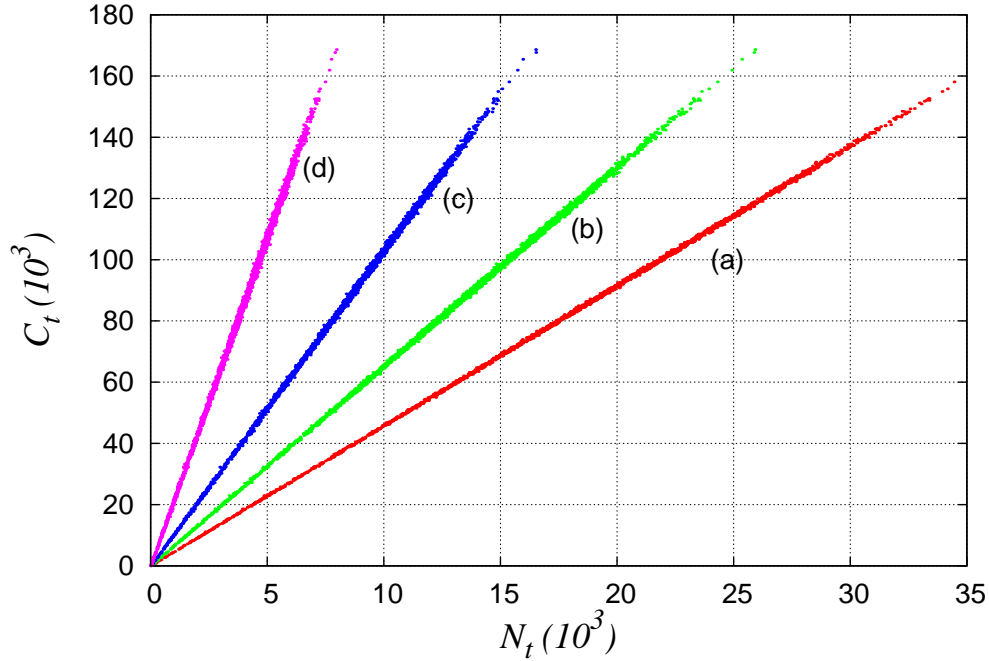}
\end{flushright}
\caption{Graphics of the algorithm complexity $C_t$, calculated by the
Zweim\"uller prescription (\ref{K}), as a function of $N_t$, the number of entrances of a given trajectory into the cell $A_1 = (x_*,1]$, during   $t=10^6$ iterations of the Bernoulli map (\ref{bern}) with $z=28/13$. For sake of clarity, only 2500 points is shown for each case (a)-(d), which correspond, respectively, to $x_*=1/2$, $5/8$, $3/4$, and $7/8$. The linear relation is evident. The situation for the other considered maps is similar.}
\label{fig2}
\end{figure}

\begin{figure}[t!]
\begin{flushright}
\includegraphics[width=0.85\linewidth]{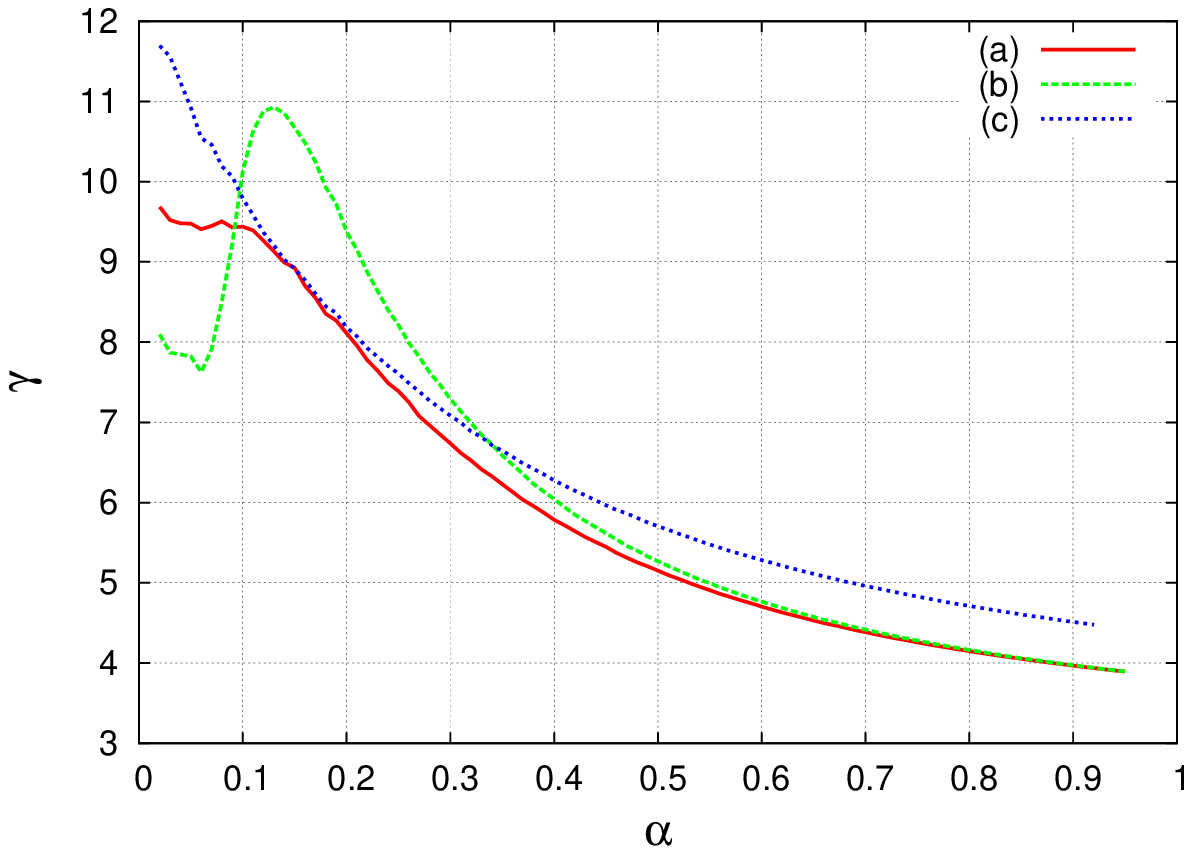}
\end{flushright}
\caption{The proportionality constant $\gamma$ between $C_t$ and $N_t$, calculated
 with respect to the standard partition, as a
function of $\alpha$ for the PM (a), Thaler (b), and Bernoulli (c) maps. For the three cases, the typical
uncertainty in $\gamma$ is about 1\% for samples of $10^4$ trajectories (computed
up to $t=10^6$). The  curves   are calculated with increments
 of $10^{-2}$ in the values of $\alpha$. }
\label{fig3}
\end{figure}

The ADK convergence of the generalized Lyapunov exponent (\ref{lamba}) was exhaustively checked and confirmed by numerical simulations for different maps in \cite{PSV}. The Zweim\"uller prescription for calculating the algorithmic complexity (\ref{K}) assures also an ADK-like convergence for the generalized KS entropy (\ref{KS}). A possible way of testing the consistence of the overall picture is to compare the Zweim\"uller prescription (\ref{K}) with other independent prescription for calculating the algorithmic complexity $C_{t}$. For this purpose, we consider some simple realizations of the general maps of the type (\ref{pmap}), namely the standard PM case discussed in \cite{GW}, the Thaler map \cite{Thaler}
\begin{eqnarray}
f(x)=x\left[1+\left(\frac{x}{1+x}\right)^{z-2}-x^{z-2}\right]^{-1/(z-2)},
\label{tmap}
\end{eqnarray}
mod 1, and the so-called modified Bernoulli map (see \cite{AAKB} for references  )
\beq
\label{bern}
f(x) = \left\{
\begin{array}{ll}
  x + 2^{z-1}x^z,  & 0 \le x \le \displaystyle \frac{1}{2}, \\
  x - 2^{z-1}(1-x)^z, & \displaystyle \frac{1}{2} < x \le 1.
\end{array}
\right.
\eeq
The modified Bernoulli map (\ref{bern}) has indeed two neutral fixed points at $x=0$ and $x=1$, but this does not alter our analysis since we still have $a_{t}\sim t^{\alpha}$ for $z>2$ in this case \cite{RZ}. Motivated by the construction introduced in \cite{GW}, let us consider the standard  partition of the interval $[0,1]$ into two cells,   $A_0 = [0,x_*]$ and   $A_1= (x_*,1]$, where $x_*$ is the point of discontinuity of the maps, {\em i.e.}, $\lim_{x\to x_*^-}f(x) = 1$, with $0<x_*<1$. For the modified Bernoulli map (\ref{bern}), one has simply $x_*=1/2$, while for the other cases the value of $x_*$ does depend on the map details, in particular on the
value of $z$. The trajectories inside both cells $A_0$ and $A_1$ are typically regular, the turbulent behavior is associated with the transition from one cell to the other, see \cite{GW} and Fig \ref{fig0}.
Let $N_t$ be the number of entrances of a given trajectory into the   cell $A_1$ during   $t$ iterations of the map. Since the contributions for $C_t$
arising from the laminar parts of the trajectories contained inside the  cells are subdominant for large $t$,
it is natural to expect that,
for weakly chaotic regimes, $C_t = \gamma N_t$ for large $t$, where $\gamma$ is some
proportionality constant, independent of the specific trajectory considered, implying, in particular, that   $C_t/\langle C_t\rangle =  N_t/\langle N_t\rangle$ for large $t$. We could check by numerical simulations that the subexponential KS entropy (\ref{KS})  calculated directly from  $N_t$, namely
\beq
\label{KS1}
 \frac{h^{(\alpha)}_t}{\langle h^{(\alpha)}_t\rangle}  =
\frac{N_t}{ \langle N_t\rangle},
\eeq
does converge toward a Mittag-Leffler distribution with unit expected value. Fig. \ref{fig1} depicts
a typical example. This convergence is robust and typically fast, see \cite{PSV} for a recent discussion. Interestingly, Eq. (\ref{KS1})  and its convergence issues  do not depend on the specific partition introduced above, even though the specific value of $\gamma$ does. Fig. \ref{fig2}
depicts the relation between the algorithm complexity calculated by using the Zweim\"uller's prescription (\ref{K}) and the values of $N_t$ for different
partitions. As one can see, both quantities  are indeed proportional, with very good accuracy, irrespective of the
partition employed. We also notice that the value of $\gamma$   depends  on the details of the maps, specifically on the value of $z$ and, consequently, of $\alpha$, see Fig. \ref{fig3}.

\section{Final remarks}

We close with some final remarks on the work \cite{KB}  and what has led   to its misleading conclusion that
\beq
\label{pes1}
h_{\mu}(f)=\alpha\left\langle \lambda^{(\alpha)}\right\rangle
\eeq
for
systems of the type (\ref{pmap}). The first observation is that (\ref{pes1}) is incompatible with the re-scaling transformation $\omega\to\xi\omega$, which must be
 a symmetry for the dynamics in the infinite measure case. The dynamical quantity on
 the right-handed side, whatever way the average is taken, must be invariant under such transformation, while the left-handed side is certainly not, see (\ref{rohlin}). Such problem can be elucidated recalling that
 the ADK theorem gives (see, for instance, \cite{PSV})
\beq
\label{ADK}
\left\langle \lambda^{(\alpha)}\right\rangle_{\rm ADK}=\frac{1}{ba}\left(\frac{a}{\alpha}\right)^{\alpha}\frac{\sin(\pi\alpha)}{\pi\alpha}\int \ln|f'(x)|\,\omega(x)dx,
\eeq
from which 
 Rohlin's formula (\ref{rohlin}) for the Krengel  entropy implies immediately that
 \begin{eqnarray}
\label{kent}
\frac{1}{b}h_{\mu}(f)= a\left(\frac{\alpha}{a}\right)^{\alpha}\frac{\pi\alpha}{\sin(\pi\alpha)}\left\langle \lambda^{(\alpha)}\right\rangle_{\rm ADK},
\end{eqnarray}
which is the correct relation between Krengel  entropy and a dynamically
meaningful  average of
subexponential Lyapunov exponents
 for maps of the type (\ref{pmap}).
 The ADK average is not only a dynamically
meaningful  average, it is essentially  {\em the} dynamically
meaningful  average for these systems. For instance, the average of the
subexponential Lyapunov exponents (\ref{lamba}) calculated for randomly chosen
(with any absolutely continuous measure with respect to the usual Lebesgue measure on the interval $[0,1]$)
 initial conditions
$x$  will converge to the ADK average for large $t$, see \cite{PSV}
for some recent applications of this important fact.
 Notice also that
 both sides of (\ref{kent}) are invariant under the symmetry  $\omega\to\xi\omega$.

 A closer inspection of \cite{KB} (see, in particular, their Eq. (10)) shows that
 they, when dealing with
 the  continuous time stochastic linear model proposed in \cite{IGR},
    tacitly choose  a value for $\xi$ such that
 \beq
b=\left(\frac{a}{\alpha}\right)^{\alpha-1}\frac{\sin(\pi\alpha)}{\pi\alpha},
\eeq
breaking the measure re-scaling symmetry of  (\ref{kent}) and rendering it in
its $\xi$-dependent form  (\ref{pes1}). However, one could have chosen any other value for $\xi$,
leading to a distinct value of $b$ and   to a completely different ``relation''
between the Krengel entropy and the ADK average. Since these relations do depend
on some specific  multiplicative constant of the infinite invariant measure, they have no dynamical meaning. It is interesting  to notice that
 $N_{t}$ is also considered as a Mittag-Leffler random variable in \cite{KB} by using renewal theory in a different manner, but its relation to $C_{t}$ is not stated.
Instead, it is used as hypothesis that $\sum_{k=0}^{t-1}\ln|f'(f^{k}(x))|\propto N_{t}$ in order to conclude that $\lambda_{t}^{(\alpha)}$ is Mittag-Leffler distributed. Such assumption presumes the convergence
\beq
\frac{\lambda_{t}^{(\alpha)}}{\left\langle\lambda_{t}^{(\alpha)}\right\rangle} \to \frac{  N_{t}}{\left\langle N_{t}\right\rangle}
\eeq
for almost all trajectories, which is indeed correct, but it is a very strong assumption without a prior  knowledge of Zweim\"uller's relation (\ref{smb}).

We close by noticing that,
 comparing (\ref{pes}) and (\ref{kent}), it is clear that the subexponential KS entropy is the appropriate entropy for extending Pesin relation for weakly chaotic systems.
Relation (\ref{pes}) is simpler than (\ref{kent}) and, mainly, it is
more powerful since it holds for almost all single trajectories, in
contrast with (\ref{kent}), where a statistic description involving many
trajectories is necessary (and, moreover, an invariant measure, which
usually is not explicitly known,  is required for the calculation of Krengel entropy).
Furthermore, the ergodic transition $\alpha \to 1$ in (\ref{kent}) is rather awkward in comparison with the same transition for the relation (\ref{pes}), which is straightforward and natural since the Mittag-Leffler distribution tends to a Dirac $\delta$-function for $\alpha\to 1$.

This work was supported by the Brazilian agencies CNPq and FAPESP.


\section*{References}
\begin{thebibliography}{99}

\bibitem{PG} P. Gaspard, {\it Chaos, Scattering and Statistical Mechanics} (Cambridge University Press, Cambridge, 1998).

\bibitem{GW} P. Gaspard, X.-J. Wang, Proc. Natl. Acad. Sci. USA {\bf 85}, 4591 (1988).

\bibitem{Aaronson} J. Aaronson, {\it An Introduction to Infinite Ergodic Theory} (American Mathematical Society, Providence, 1997).

\bibitem{Pesin} Ya. B. Pesin, Russ. Math. Surveys {\bf 32}, 55 (1977).

\bibitem{KB} N. Korabel and E. Barkai, Phys. Rev. Lett. {\bf 102}, 050601 (2009); Phys. Rev. E {\bf 82}, 016209 (2010).

\bibitem{GTA} G.F.J. A\~na\~nos and C. Tsallis, Phys. Rev. Lett. {\bf 93}, 020601 (2004); P. Grassberger, Phys. Rev. Lett. {\bf 95}, 140601 (2005).

\bibitem{PM} Y. Pomeau and P. Manneville, Commun. Math. Phys. {\bf 74}, 189 (1980).

\bibitem{RZ} R. Zweim\"uller, Ergod. Th. \& Dynam. Sys. {\bf 20}, 1519 (2000).



\bibitem{Thaler} M. Thaler, Studia Math. {\bf 143}, 103 (2001).

\bibitem{AAC} T. Akimoto and Y. Aizawa, Chaos {\bf 20}, 033110 (2010).

\bibitem{PSV} C.J.A. Pires, A. Saa, and R. Venegeroles,
 Phys. Rev. E. {\bf 84} 066210, (2011).


\bibitem{BWZ} R. Zweim\"uller, Discrete Contin. Dyn. Syst. {\bf 15}, 353 (2006).

\bibitem{Kr} U. Krengel, Z. Wahrscheinlichkeitstheor. Verw. Geb. {\bf 7}, 161 (1967).

\bibitem{Rohlin} V.A. Rohlin, Am. Math. Soc. Transl. II ser. {\bf 39}, 1 (1964).

\bibitem{Feller} W. Feller, {\it An Introduction to Probability Theory and its Applications - Vol. II} (Wiley, New York, 1971).

\bibitem{SV} A. Saa and R. Venegeroles, Phys. Rev. E {\bf 84}, 026702 (2011).

\bibitem{AAKB} T. Akimoto and Y. Aizawa, J. Korean Phys. Soc. {\bf 50}, 254 (2007).

\bibitem{IGR} M. Ignaccolo, P. Grigolini, and A. Rosa, Phys. Rev. E {\bf 64}, 026210 (2001).





\end{thebibliography}
\end{document}